\documentclass[nofootinbib,reprint,amsmath,amssymb,prd,aps,superscriptaddress]{revtex4-1}
\usepackage{graphicx}
\usepackage[T1]{fontenc}
\usepackage{dcolumn}
\usepackage{xcolor,colortbl}
\usepackage{multirow}
\usepackage{comment}
\usepackage{enumitem}
\usepackage[caption=false]{subfig}
\usepackage{float}
\usepackage{bm}
\newcommand{\qm}[1]{``#1''}
\usepackage{hyperref}
\hypersetup{colorlinks, linkcolor={red},citecolor={blue},urlcolor={blue}}   

\newcommand{\dd}[0]{{\rm d}}

\DeclareUnicodeCharacter{2212}{-}

\begin{document}

\title[Can wormholes mirror the quasi-normal mode spectrum of Schwarzschild black holes?]{Can wormholes mirror the quasi-normal mode spectrum\\ of Schwarzschild black holes?}

\author{Ciro~De~Simone}
\email{ciro.desimone@unina.it}
\affiliation{Dipartimento di Fisica \qm{E. Pancini}, Universit\'a di Napoli \qm{Federico II}, Complesso Universitario di Monte S. Angelo, Via Cintia Edificio 6, I-80126 Napoli, Italy}
\affiliation{Istituto Nazionale di Fisica Nucleare, Sezione di Napoli, Complesso Universitario di Monte S. Angelo, Via Cintia Edificio 6, 80126 Napoli, Italy}

\author{Vittorio~De~Falco}
\email{v.defalco@ssmeridionale.it}
\affiliation{Scuola Superiore Meridionale,  Largo San Marcellino 10, 80138 Napoli, Italy}
\affiliation{Istituto Nazionale di Fisica Nucleare, Sezione di Napoli, Complesso Universitario di Monte S. Angelo, Via Cintia Edificio 6, 80126 Napoli, Italy}

\author{Salvatore~Capozziello}
\email{capozziello@na.infn.it}
\affiliation{Dipartimento di Fisica \qm{E. Pancini}, Universit\'a di Napoli \qm{Federico II}, Complesso Universitario di Monte S. Angelo, Via Cintia Edificio 6, I-80126 Napoli, Italy}
\affiliation{Scuola Superiore Meridionale,  Largo San Marcellino 10, 80138 Napoli, Italy}
\affiliation{Istituto Nazionale di Fisica Nucleare, Sezione di Napoli, Complesso Universitario di Monte S. Angelo, Via Cintia Edificio 6, 80126 Napoli, Italy}

\date{\today}

\begin{abstract}
Wormholes are exotic compact objects characterized by the absence of essential singularities and horizons, acting as slender bridges linking two distinct regions of spacetime. Despite their theoretical significance, they remain however undetected, possibly due to their ability to closely mimic the observational properties of black holes. This study explores whether a static and spherically symmetric wormhole within General Relativity can reproduce the quasi-normal mode spectrum of a Schwarzschild black hole under scalar, electromagnetic, and axial gravitational perturbations, both individually and in combination. To address this, we reformulate the wormhole metric components using a near-throat parametrization. Our analysis concentrates on the fundamental mode and first overtone, estimated via the Wentzel–Kramers–Brillouin method. By employing a customized minimization strategy, we demonstrate that within a specific region of the parameter space, a wormhole can successfully replicate a subset of the black hole quasi-normal mode spectrum.
\end{abstract}

\maketitle

\section{Introduction}
\label{sec:intro}
The first direct detection of gravitational waves (GWs) from merging of two black holes (BHs), announced by LIGO and Virgo collaborations in 2015, revolutionized modern physics \cite{LIGOScientific:150914}. Indeed, since then several other events ($\gtrsim 100$\footnote{See \href{https://gwosc.org/eventapi/html/allevents/}{https://gwosc.org/eventapi/html/allevents/}.}) have been observed, including not only binary BHs, but also collisions of two neutron stars (NSs) \cite{Abbott2023}. GW astronomy offers new insights into the universe across both the smallest and largest astrophysical scales, opening thus an additional research window on the investigation of gravity and physics of compact objects \cite{Bailes2021}.

BHs and NSs have been extensively studied in X-ray astronomy \cite{Psaltis2008,Kalemci2018,EHT}; however, methods within this domain are less effective than those in GW science for probing regions beneath the photonsphere, where new physics may emerge \cite{Colleoni2024}. Furthermore, \emph{exotic compact objects} —either beyond General Relativity (GR) predictions or arising from unconventional GR assumptions— could theoretically exist, though they remain undetected \cite{Bezares2024}. This elusiveness may be due to their ability to closely mimic the observational properties of BHs. Any distinguishable feature is expected to appear primarily in the strong-field regime (still partially explored), highlighting thus the privileged value of GW theory for testing gravity within GR and alternative frameworks \cite{Giddings2019}. On the other hand, some exotic compact objects, in the framework of GR extensions, may have  already been detected \cite{Astashenok:2020qds, Astashenok:2021peo},  but their final  nature is far to be completely defined.

One of the methods to discriminate among different compact objects consists of resorting to the quasi-normal modes (QNMs). \emph{They are distinctive vibrations of spacetime, arousing in response to disturbances} \cite{Chandrasekhar1998book,Berti:2009kk}. When a compact object is perturbed (e.g., by the collapse of a star or the merger of dense objects), it undergoes temporary reverberations, before returning to a stable state. These undulations are termed \emph{quasi-normal}, because they do not continue indefinitely, but they are gradually damped by the GW emission, which leads then to a steady decay over time. Mathematically, each QNM is represented by a complex frequency: the real part sets the oscillation rate, while the imaginary part defines the damping level, indicating how swiftly the resonance fades. The QNM spectrum consists of an infinite set of QNM frequencies, each one identified by an integer $n$, referred to as the \emph{overtone number} in literature\footnote{For completeness, a QNM frequency is identified by three integer indices $n,l,s$, where the last two will be introduced later.}. The QNM spectrum provides valuable information on the properties of an astrophysical object \cite{Ferrari2008,Salcedo2019}. For instance, the QNMs of a Kerr BH are fully specified by mass and angular momentum parameters \cite{Chandrasekhar1998book}. 

QNMs are mainly excited during the merger of compact objects, dominating the ringdown phase, the final stage of the merger \cite{Destounis2023}. Current measurements of ringdown signals and QNMs carry significant uncertainties, making it challenging to definitively identify the nature of the source \cite{Toubiana2020}. To date, observations agree with the BH model predicted by GR \cite{Nakano2019}, yet they do not exclude possible alternative sources, such as wormholes (WHs).

The latter exotic compact objects can be regarded as peculiar extensions of BHs, with the added stipulation that spacetime remains regular throughout. They feature two asymptotically flat regions joined by a narrow bridge, or throat, devoid of event horizons and essential singularities \cite{Visser1995}. To ensure stability and traversability within the GR framework, WHs necessitate the presence of \emph{exotic matter}, being the set of mechanisms violating the conventional energy conditions \cite{Morris1988}.

The studies on WHs can be broadly classified into two principal areas: (1) proposing new WH solutions within the framework of GR \cite{Bronnikov2013,Garattini2019,Konoplya2022,Armendariz2002} or alternative theories of gravity, employing either exotic stress-energy tensors \cite{Ebrahimi2010,Bohmer2012,Harko2013,Bahamonde2016,Mustafa2021,Battista:2024gud}, purely gravitational topological constructs \cite{Defalco2023WHpureg} or matter fields satisfying the energy conditions \cite{Eiroa2008,Lobo2009,Capozziello2012, Capozziello:2022zoz}; (2) developing astrophysical methods to identify observational signatures of WHs, leveraging techniques in X-ray domain \cite{Ohgami2015,Paul2019,Defalco2020WH,DeFalco2021,DeFalco2021WF,DeFalco2021EF,Defalco2023} or GW astronomy \cite{Hong2021,Konoplya2016,Cardoso2016}. 

Numerous studies explored QNMs related to WHs, with particular emphasis on the similarities between the ringdown phase of a WH and that of a BH \cite{DamourWH, KonoplyaWH, Konoplya2005, Cardoso2016}. This research line often involves comparing the QNMs of BHs with those of specific WH models (e.g., Ellis-Bronnikov WH \cite{KonoplyaWH}, Einstein-Rosen solution \cite{Cardoso2016}), predominantly focusing on scalar-type perturbations \cite{KonoplyaWH, DamourWH}. 

This paper aims to advance this research area by broadening its scope along four significant directions: (1) it examines a more extensive class of WHs in GR, namely general static and spherically symmetric spacetimes; (2) it includes perturbations of various spins: scalar, electromagnetic, and gravitational; (3) it adopts a parametrization designed to encapsulate the characteristics of a WH spacetime near its throat; (4) it employs a robust quantitative approach, utilizing the Wentzel–Kramers–Brillouin (WKB) approximation to calculate QNM frequencies, with particular emphasis on fundamental mode\footnote{A \emph{mode} refers to a specific solution of the perturbation equations describing oscillations in the system and it is characterized, as we will see, by three different indices, where $n$ is included.} ($n=0$) and lower overtones due to their observational significance. \emph{Our primary goal is to assess how well a WH in GR can replicate the characteristic spectrum of a Schwarzschild BH}.

The structure of the paper is as follows: in Sec. \ref{sec:II}, we outline the main features of a Morris-Thorne-like WH and introduce the near-throat parametrization; Sec. \ref{sec:III} is dedicated to examining various perturbations over a WH background; in Sec. \ref{sec:IV}, we describe the method used to estimate the QNM frequencies and to compare the related WH spectra with that of the Schwarzschild BH; the results of this analysis are presented in Sec. \ref{sec:V}, while Sec. \ref{sec:VI} concludes with a summary of our findings and potential directions for future research.

\section{Near-throat parametrization of static and spherically symmetric wormholes}
\label{sec:II}
We present the general features of a static and spherically symmetric WH described by the Morris-Thorne-like metric (see Sec. \ref{sec:IIA}) and the near-throat parametrization of the metric coefficients (see Sec. \ref{sec:IIB}). Throughout the paper we adopt geometrical units: $G=c=1$.

\subsection{The Morris-Thorne-like metric of a static and spherically symmetric wormhole}
\label{sec:IIA}
The Morris-Thorne-like metric describes a static and spherically symmetric WH, whose line element can be expressed in spherical-like coordinates $(t,r,\theta,\varphi)$ as \cite{Morris1988}
\begin{equation}\label{MT metric}
\dd s^2=-e^{2\Phi(r)} \dd t^2+\frac{\dd r^2}{1-\frac{b(r)}{r}}+r^2(\dd\theta^2+\sin^2\theta \dd\varphi^2).
\end{equation}
Such WHs are characterized by two key elements: the \emph{redshift function} $\Phi(r)$, representing the gravitational redshift experienced by light or signals traveling through it; the \emph{shape profile} $b(r)$, defining its spatial structure.

Notice that the coordinate $r$ extends to $r=\infty$ in the two spacetime regions connected by the WH throat, corresponding to the minimum value $r=b_0$. Due to the spherical symmetry hypothesis, we can set without loss of generality $\theta=\pi/2$ in Eq. \eqref{MT metric}. The functions $\Phi(r)$ and $b(r)$ have to satisfy a series of properties:
\begin{itemize}
    \item \emph{asymptotic flatness}, namely we must have
    \begin{equation}
     \lim_{r \to \pm\infty}\Phi(r)=0,\qquad \lim_{r \to \pm\infty}\frac{b(r)}{r}=0;   
    \end{equation}
    \item \emph{absence of horizons and essential singularities}: $\Phi(r)$ and $b(r)$ are smooth and finite functions in the domain of definition $\mathcal{D}=[-\infty,-b_0]\cup[b_0,+\infty]$;
    \item \emph{appropriate profiles}: given $r_1<r_2$ in a universe, we have $g_{tt}(r_1)>g_{tt}(r_2)$ and $g_{rr}(r_1)>g_{rr}(r_2)$, which implies $\Phi(r_1)<\Phi(r_2)$ and $b(r_1)/r_1>b(r_2)/r_2$, namely $\Phi(r)$ and $b(r)$ are monotonic increasing and decreasing functions, respectively;
    \item \emph{introduction of a proper radial distance} from a point at radial coordinate $r$ to the WH throat can be constructed by requiring that $b(r)\le r$ and $b(b_0)=b_0$. In these hypotheses, we can define:
    \begin{equation}
    l(r)=\pm \int_{b_0}^\infty \frac{\dd r^\star}{\sqrt{1-\frac{b(r^\star)}{r^\star}}};    
    \end{equation}
    \item \emph{flaring out condition} is a geometric requirement ensuring the WH throat remains open and traversable. It guarantees that the WH geometry expands out as it moves away from the throat, rather than constricting or collapsing \cite{Morris1988}. Mathematically, it is expressed by $b'(r) < 1$ at $r=b_0$, where the prime denotes the derivative with respect to $r$. It can be shown this prescription can only be satisfied within GR by considering a stress-energy tensor $T_{\mu\nu}$ that violates the energy conditions, implying thus the need for exotic matter; 
    \item \emph{parameter's dependence} sees as first the presence of the Arnowitt, Deser, Misner (ADM) mass $M$, defined as the total mass-energy of the system contained in the whole spacetime \cite{Visser1995}, i.e., 
\begin{equation} \label{eq:ADMmass}
M= \lim_{r\to+\infty}m(r)=\frac{b_0}{2}+4\pi\int^{\infty}_{b_0}\rho(x)x^2 dx,
\end{equation}
where $\rho(r)$ is the mass-energy density. Based on the underlying gravity framework, a WH solution, besides the ADM mass, can reckon also on other parameters (see e.g, Refs. \cite{DeFalco2021,DeFalco2021EF,Defalco2023}).     
\end{itemize} 

\subsection{Near-throat parametrization}
\label{sec:IIB}
For the objectives of this work, we employ the \emph{near-throat parametrization} \cite{Konoplya2018, Chianese2017}. It assumes that the redshift and shape functions can be expressed as Taylor-series expansions near the WH throat $b_0$ as follows
\begin{align}\label{eq:NT-parametrization}
\Phi(r)&=\sum_{i=0}^\infty \Phi_i(r-b_0)^i,\quad b(r)=\sum_{i=0}^\infty b_i(r-b_0)^i,
\end{align}
where $\{\Phi_i\}_{i=0}^\infty$ and $\{b_i\}_{i=0}^\infty$ are sets of real constant coefficients. This parametrization is justified by the observation that the QNM spectrum of a compact object is mostly determined by the local form of the potential near the gravitational source (where it attains its maximum), rather than by its asymptotic properties \cite{Konoplya2011}.

It is important to note that applying this parametrization to the metric \eqref{MT metric} results in a spacetime that is no longer asymptotically flat. This issue can be resolved by introducing suitable \emph{pre-factors} $f_j(r)$, $j=1,2,3,4$, being 
\begin{subequations}
    \begin{align}
        g_{tt}&= -f_1(r)\,e^{2\Phi(r)}+f_2(r),\\
        g_{rr}&=\frac{f_3(r)}{1-\frac{b(r)}{r}}+f_4(r).
    \end{align}
\end{subequations}
They must fulfill two key conditions: 
\begin{itemize}
    \item[(1)] they must reduce to unity near the WH throat, namely for $r \to b_0$, we have $f_2(r)\to 0$ and $f_4(r)\to 0$, whereas $f_1(r)\to 1$ and $f_3(r)\to1$;
    \item[(2)] they must enforce the correct asymptotic flatness behavior of the metric at large radial distances, namely for $r \to \infty$, we have $f_1(r)e^{2\Phi(r)}\to0$, $f_2(r)\to-1$, $f_3(r)/(1-b(r)/r)\to 0$, and $f_4(r)\to 1$.
\end{itemize} 
It is thus clear that specifying the behavior of the WH metric near the throat and at infinity does not uniquely determine the form of the pre-factors. This outcome is unsurprising, as the near-throat parametrization is independent of the metric's behavior at large distances. As a consequence, it is in principle possible to restrict the class of pre-factors to those that leave the near-throat metric unaltered. In this way the WH QNM spectrum is fully captured by the near-throat parametrization.

\section{Perturbations over a wormhole background}
\label{sec:III}
In this section, we first provide the preliminary hypotheses behind our analysis (see Sec. \ref{sec:preliminary}) for then studying these perturbations over the WH background \eqref{MT metric}: scalar (see Sec. \ref{sec:IIIA}), electromagnetic (see Sec. \ref{sec:IIIB}), and gravitational of axial parity (see Sec. \ref{sec:IIIC}). 

\subsection{Preliminary hypotheses}
\label{sec:preliminary}
For the purposes of this paper, it is useful to introduce the (generalized) tortoise coordinate $r_*$ defined as \cite{del-Corral2022}
    \begin{equation}\label{eq:tortoise_coordinate}
\dd r_*=\sqrt{\frac{g_{rr}}{g_{tt}}}\dd r\ \ \Rightarrow\ \      r_*=\int \frac{e^{-\Phi(r)}}{\sqrt{1-\frac{b(r)}{r}}}\dd r,
    \end{equation}
where at $r_* = \pm\infty$ correspond to the spatial infinity of the two universes. Since there is no reflecting potential, the WH throat plays the role of the BH event horizon, adopting thus the same methods as the BH case \cite{Konoplya2005}. For an arbitrary choice of $\Phi(r)$ and $b(r)$ the integral \eqref{eq:tortoise_coordinate} may not be solvable analytically. However, this does not affect the WKB method, which compute $\frac{\dd V}{\dd r_*}=\frac{\dd r}{\dd r_*} \frac{\dd V}{\dd r}$.

Similarly to what happens to BHs, variations of a static and spherically symmetric WH satisfy a \emph{time-independent Schrödinger-like equation} in the (generalized) tortoise coordinate, with a radial potential $V(r)$ strongly depending on the perturbation-type \cite{Kim2008}:
    \begin{equation}\label{eq:SEgeneral}
        \frac{\dd^2 \Psi^{(i)}(r)}{\dd r_*^2}+\Big[\omega^2-V^{(i)}(r)\Big]\Psi^{(i)}(r)=0,
    \end{equation}
where the index $(i)$ over a quantity denotes the spin-$i$ of the perturbation, where $i=0,1,2$ for scalar, electromagnetic, and gravitational perturbations, respectively. 

It is relevant to stress that, although the perturbation equation has a similar form for BH and WH, the tortoise coordinate is defined differently. \emph{This implies that any resemblance between the perturbation potentials does not necessarily reflect in the ensuing QNM spectrum.}

The QNMs of BHs are characterized by the following \emph{boundary conditions} \cite{Konoplya2005}: ingoing radiation into the BH event horizon and outgoing radiation at infinity. As previously observed, the similarities existing between BHs and WHs permit to define these boundary conditions for WHs: outgoing radiation in the two asymptotically flat spacetimes and no radiation coming into the WH from either side, namely $\Psi_l \sim e^{\pm i \omega r_*}$ as $r_* \to \pm \infty$.

Our analysis disregards also contributions from \emph{GW echoes}. These are hypothetical signals that would be emitted by a perturbed horizonless compact object endowed with a photonsphere \cite{Pani2019}, arising from the presence of multiple peaks in the potential $V^{(i)}(r)$ \cite{Volkel2018}. By neglecting echoes, we effectively simplify the problem, limiting the class of potentials to those with a single peak. 

\subsection{Scalar perturbations}
\label{sec:IIIA}
The evolution of a massless scalar field $\eta$ over the background \eqref{MT metric} is ruled by the Klein-Gordon equation \cite{Konoplya2011}:
    \begin{equation}\label{KG}
         \Box \,\eta \equiv\frac{1}{\sqrt{-g}}\partial_\mu (\sqrt{-g}\, g^{\mu\nu}\partial_\nu \eta)=0,
    \end{equation}
where $\Box=g^{\mu\nu}\partial_\mu\partial_\nu$ is the relativistic d'Alembert operator. We disregard any scalar field's back-reaction effect on the background spacetime. Equation \eqref{KG} relies solely on the form of the metric $g_{\mu\nu}$, as it does not depend on the field equations of the underlying gravity theory.     

Since the background is spherically symmetric, we can separate the solution 
$\eta(t,r,\theta,\varphi)$ into time, radial, and angular parts. We first start from the time component, assuming the \emph{harmonic dependence}, which is given by:
\begin{equation}
\eta(t,r,\theta,\varphi)=e^{-i\omega t}\psi(r,\theta,\varphi),    
\end{equation}
where $\omega$ is the so-called \emph{QNM frequency}.

Invoking the spherical symmetry's hypothesis, the spatial part $\psi(r,\theta,\varphi)$ can be expanded in terms of the spherical harmonics $Y^{lm}(\theta,\varphi)$ as follows \cite{Konoplya2011}:
    \begin{equation}         \psi(r,\theta,\varphi)=\sum_{l=0}^{\infty}\sum_{m=-l}^{l}\frac{\Psi^{(0)}_{lm}(r)}{r}Y^{lm}(\theta,\varphi),
    \end{equation}
where the sum extends over the \emph{multipole index} $l$ and the \emph{azimuthal number} $m$ of the spherical harmonics. Substituting this ansatz into Eq. \eqref{KG} and using the spherical harmonics' properties, we can separate the radial part $\Psi^{(0)}_{lm}(r)$, which satisfies the following Schr{\"o}dinger-like equation (in spherical symmetry there is no dependence on the index $m$ and so will be suppressed) \cite{Konoplya2011}:
    \begin{equation}\label{eq:SE}
        \frac{\dd^2 \Psi^{(0)}_{l}(r)}{\dd r_*^2}+\Big[\omega^2-V^{(0)}_l(r)\Big]\Psi^{(0)}_{l}(r)=0,
    \end{equation}
where the potential $V^{(0)}_l(r)$ reads as
\begin{equation} \label{eq:POT_scalar}
V^{(0)}_l(r)=e^{2\Phi(r)}\frac{l(l+1)}{r^2}+\frac{1}{2r}\frac{\dd}{\dd r} \left[e^{2\Phi(r)}\left(1-\frac{b(r)}{r}\right)\right].
\end{equation}

\subsection{Electromagnetic perturbations}   
\label{sec:IIIB}
To study massless electromagnetic (vector) perturbations $F_{\mu\nu}$ in absence of sources, we must deal with the Maxwell equations in the curved spacetime \eqref{MT metric} \cite{Konoplya2011}
\begin{equation}\label{eq:Maxwell}
\nabla_\nu F^{\mu\nu}\equiv\frac{1}{\sqrt{-g}}\partial_\nu(\sqrt{-g} F^{\mu\nu})=0,
\end{equation}
where $F_{\mu\nu}=\nabla_\nu A_{\mu}-\nabla_\mu A_{\nu}=\partial_\nu A_{\mu}-\partial_\mu A_{\nu}$ is an antisymmetric tensor, which can be written in terms of the electromagnetic four potential $A_\mu$.

Imposing the Bianchi identities (including no magnetic monopoles), we obtain this differential equation \cite{Konoplya2011}:
\begin{equation}\label{eq:Bianchi}
\partial_\lambda F_{\mu\nu}+\partial_\mu F_{\nu\lambda}+\partial_\nu F_{\lambda\mu}=0.
\end{equation}

These two sets of differential Eqs. \eqref{eq:Maxwell} and \eqref{eq:Bianchi} for $F_{\mu\nu}$ can be decomposed into scalar and vector spherical harmonics to separate variables in $t, r,\theta,\varphi$ coordinates. $F_{\mu\nu}$ can be split into electric (polar or even-parity) and magnetic (axial or odd-parity) modes, generally leading to a different type of perturbation equation \cite{Nollert1999}. It is possible to prove then that in spherically symmetric spacetimes, the isospectrality between polar and axial modes always holds \cite{Cardoso2019}. Therefore, the radial component $\Psi^{(1)}_{l}(r)$ satisfies Eq. \eqref{eq:SE} with the following unique form \cite{Cardoso2019}
\begin{equation}\label{eq:POT_vector}
V^{(1)}_l=e^{2\Phi(r)}\frac{l(l+1)}{r^2}.
\end{equation}
Notice that this potential only depends on the time component of the WH metric and has a much simpler expression than the potential for scalar perturbations \eqref{eq:POT_scalar}. Also in this case, Eq. \eqref{eq:POT_vector} does not depend on the field equations of the underlying gravity theory.

\subsection{Axial gravitational perturbations}
\label{sec:IIIC}
A gravitational perturbation $h_{\mu\nu}$ over the spherically symmetric spacetime \eqref{MT metric}, $g^{(0)}_{\mu\nu}$, can be written as \cite{Konoplya2011}
\begin{equation}
g_{\mu\nu}=g^{(0)}_{\mu\nu}+h_{\mu\nu},
\end{equation}
where $g_{\mu\nu}$ is the perturbed metric. Notice that $h_{\mu\nu}$ does not have to be spherically symmetric, entailing thus the metric $g_{\mu\nu}$ has no particular symmetries. From the spherical symmetry hypothesis, $h_{\mu\nu}$ can be decomposed in terms of scalar, vector, and tensor spherical harmonics \cite{Pani2013}. Those functions have well-defined properties under parity transformations, which allow to separate gravitational perturbations in two main classes: axial (or odd) and polar (or even) parity. Axial and polar gravitational perturbations should be labeled as $s=2-$ and $s=2+$, respectively. However, since we deal only with the former, we lighten the notations just writing $s=2$.

Axial perturbations cause the spacetime to rotate and change sign when the azimuthal angle is reversed, $\varphi \to \varphi + \pi$, indeed under parity transformation they gain a factor of $(-1)^{l+1}$. In contrast, polar perturbations do not produce any rotation and transform with a factor of $(-1)^{l}$ under the parity map \cite{Pani2013}. Both axial and polar perturbations satisfy a Schr{\"o}dinger-like equation, similarly to what happens in the scalar and electromagnetic cases. For the purposes of this paper we will only be interested in axial gravitational perturbations, since they decouple from the perturbations of the energy-momentum tensor. This is not the case, in general, for polar perturbations, which require a separate and more intricate treatment.

The potential for axial perturbations is given by \cite{Kim2008}
\begin{align}\label{eq:potential_axial}
V_l^{(2), \rm axial}(r) &= \frac{e^{2\Phi(r)}}{r^3}\Biggl[r(l+2)(l-1)-\Phi'(r)\Delta(r)\notag\\ 
& -\frac{1}{2}\Delta'(r)+\frac{3\Delta(r)}{r}\Biggr],
\end{align} where $\Delta(r)=r^2-b(r)r$. More details on the derivation of the potential \eqref{eq:potential_axial} can be found in Appendix \ref{AppendixA}. In this case instead, Eq. \eqref{eq:potential_axial} depends on the field equations.
    
\section{Methodology}
\label{sec:IV}
This section focuses on the methodology used in our work. We outline the key aspects of the WKB approximation to compute the QNMs (see to Sec. \ref{sec:IVA}) and we describe the technique able to search for the isospectrality between BHs and WHs in GR (see Sec. \ref{sec:IVB}). 

\subsection{The WKB method}
\label{sec:IVA}
Across the range of parameters defined in Eq. \eqref{eq:NT-parametrization}, the potentials associated with scalar, electromagnetic, and gravitational WH perturbations exhibit two remarkable characteristics: they converge to a constant value at large distances and feature a single peak at the WH throat, $b_0$. These properties represent the essential requirements for applying the WKB method to calculate QNMs \cite{IyerWKB}. This approach yields analytical approximate expressions for the QNMs and has been effectively employed in different BH (e.g., \cite{IyerWKB,Kokkotas:1988fm,Seidel:1989bp}) and WH spacetimes (e.g., \cite{Kim2008}). The WKB method achieves its highest accuracy when $n\leq l$, making it well-suited for studying the fundamental mode and the lowest overtones. Its two primary strengths are high precision and computational efficiency, making it helpful for systematic BH-WH comparisons.

This method envisages that the Schr{\"o}dinger-like equation \eqref{eq:SEgeneral} is solved separately in three domains (see Ref. \cite{IyerWKB}, for more details): the two asymptotic regions, where the potential approaches a constant value, and near the WH throat, where the potential is expressed as a Taylor series around $b_0$. The WKB approximation's order is determined by the truncation level of the Taylor expansion, with the $n$-th order corresponding to a potential expanded up to $2n$-th order. Furthermore, the WKB method has been extended to sixth order \cite{Konoplya2003} and also up to thirteenth order via the Padé approximants \cite{Matyjasek2017}.

For the aim of this paper, it is enough to employ the WKB method 
at the third order \cite{IyerWKB}. The explicit formula for the QNM frequencies reads as 
\begin{align}
\omega^2 &= \left[V_0+\sqrt{-2\ddot{V}_0} \Lambda(n)\right]-i \alpha \sqrt{-2\ddot{V}_0} \Big[1+\Omega(n)\Big],
\end{align}
where the over dot denotes the derivative with respect to the generalized tortoise coordinate, $V_0$ and $\ddot{V}_0$ represent the potential and its second derivative evaluated both at the peak value, and $\alpha=n+1/2$. The functions $\Lambda(n)$ and $\Omega(n)$ depend only on $n$ and derivatives of the potential evaluated at the peak up to the sixth order\footnote{The explicit expressions of $\Lambda(n)$ and $\Omega(n)$ can be found in Eqs. (1.5a) and (1.5b) of Ref. \cite{IyerWKB}, respectively.}. The dependence on the multipole index $l$ is encapsulated in the potential $V_0$ and its derivatives. In the BH case, the QNM frequency $\omega$ is normalized with respect to the BH mass, whereas for WHs is done with respect to the throat $b_0$, being related to the WH mass (see Eq. (\ref{eq:ADMmass})). This implies we can rescale the perturbation potentials introduced in Sec. \ref{sec:III} in unity of $b_0$, namely as $r/b_0$.

\subsection{Comparison strategy}
\label{sec:IVB}
We first introduce a notation to distinguish between WH and BH real and imaginary parts of the QNMs, as well as modes of different overtones $n$, multipole indices $l$, and spin-perturbation $s$. To this end, a general QNM will be identified by a triad of indices $(n,l,s)$ as follows: $\omega^{\rm BH}_{\rm nls}$ for BHs and $\omega^{\rm WH}_{\rm nls}$ for WHs, where the latter frequencies are functions of the parameters (cf. Eq. (\ref{eq:NT-parametrization}))
\begin{equation}\label{eq:PARAMETERS}
\zeta_N=\{b_0,b_1,\cdots,b_N,\Phi_0,\Phi_1,\cdots,\Phi_N\}.    
\end{equation}

The Schwarzschild BH's QNM spectrum is reported in Tab. \ref{tab: reference frequencies}, as a reference for the subsequent comparisons. The selected indices fulfill the condition $n\leq l$ of the WKB method and will be applied also to the WH case.
\renewcommand{\arraystretch}{1.8}
\begin{table}[!ht]
\centering
\caption{Scalar, electromagnetic, and gravitational QNM frequencies $\omega_{\rm nls}^{\rm BH}$ pertaining to the Schwarzschild BH in units of mass $M$, and decomposed in real and imaginary parts.\vspace{0.2cm}}
\begin{tabular}{|c|c|c|} 
\hline
$\qquad\boldsymbol{\omega^{\rm BH}_{\rm nls}}\qquad$ & $\qquad\boldsymbol{\Re \,[ \omega^{\rm BH}_{\rm nls}]}\qquad$ & $\qquad\boldsymbol{\Im \,[ \omega^{\rm BH}_{\rm nls}]}\qquad$\\
\hline
\multicolumn{3}{|c|}{\bf Scalar} \\
\hline
$\omega^{\rm BH}_{000}$ & 0.1046 & -0.1152\\
\hline
$\omega^{\rm BH}_{010}$ & 0.2911 & -0.0980 \\
\hline
\multicolumn{3}{|c|}{\bf Electromagnetic} \\ 
\hline
$\omega^{\rm BH}_{011}$ & 0.2459 &-0.0931 \\
\hline
$\omega^{\rm BH}_{111}$ & 0.2113& -0.2958\\
\hline
\multicolumn{3}{|c|}{\bf Gravitational (axial)} \\
\hline
$\omega^{\rm BH}_{022}$ & 0.3732&-0.0892\\
\hline
$\omega^{\rm BH}_{122}$ & 0.3460&-0.2749\\
\hline
\end{tabular}
\label{tab: reference frequencies}
\end{table}

For quantifying the comparisons, we define the relative error between BH and WH QNM frequencies for the real $\delta \omega^{\rm (r)}_{\rm nls}$ and imaginary $\delta \omega^{\rm (i)}_{\rm nls}$ components as follows:
\begin{subequations}
\begin{align}
\delta \omega^{\rm (r)}_{\rm nls}(\zeta_N) &= \frac{\Re \,[ \omega^{\rm BH}_{\rm nls}-\omega^{\rm WH}_{\rm nls}(\zeta_N)]}{\Re [\omega^{\rm BH}_{\rm nls}]} \label{rel_error_real},\\
\delta \omega^{\rm (i)}_{\rm nls}(\zeta_N) &= \frac{\Im \,[  \omega^{\rm BH}_{\rm nls}-\omega^{\rm WH}_{\rm nls}(\zeta_N)]}{\Im [\omega^{\rm BH}_{\rm nls}]}, \label{rel_error_imag}
\end{align}
\end{subequations}
where $\Re[\cdot]$ and $\Im[\cdot]$ stay for real and imaginary part of a function enclosed between square brackets, respectively. 

The combined relative error $\delta \omega^{\rm (c)}_{\rm nls}$, including the two previous measurements into a single one, reads as \cite{Volkel2019, del-Corral2022}: 
    \begin{equation}
         \delta \omega^{\rm (c)}_{\rm nls}(\zeta_N) = \sqrt{\left[\delta \omega^{\rm (r)}_{\rm nls}(\zeta_N)\right]^2+\left[\delta \omega^{\rm (i)}_{\rm nls}(\zeta_N)\right]^2}.
    \end{equation}
    
To account for various $\delta \omega^{\rm (c)}_{\rm nls}(\zeta_N)$ with distinct $l$ and $n$ and fixed spin $s$, we introduce the following function:
\begin{equation}
        \Upsilon_{\rm nls}(\zeta_N)=\sum_{i=0}^{\rm n} \sum_{j=s}^{\rm l} \delta\omega_{ijs}^{(c)}(\zeta_N).
\end{equation}

Since we are interested in comparing the aforementioned modes also for all quoted spin perturbations, we introduce the following last function:
    \begin{equation}
        \bar\Upsilon(\zeta_N) = \delta \omega^{\rm (c)}_{\rm 000}(\zeta_N)+\delta \omega^{\rm (c)}_{\rm 011}(\zeta_N)+\delta \omega^{\rm (c)}_{\rm 022}(\zeta_N),
    \end{equation}
which can be easily generalized to take into account QNM combinations endowed with different $n$, $l$, and $s$.

To effectively determine the isospectrality between BHs and WHs in GR, we need to find the minimum value of $\Upsilon_{\rm nls}(\zeta_N)$ or $\bar\Upsilon(\zeta_N)$ in terms of the parameter set $\zeta_N$, which allows to reconstruct the WH solution/s.

Minimizing the function $\Upsilon_{\rm nls}(\zeta_N)$ (or $\bar{\Upsilon}(\zeta_N)$) alone does not suffice to identify the best model parameters; it is also necessary to ensure the potential satisfies the WKB approximation criteria within the chosen area of the parameter space. This requires systematic sampling of the target region and examination of the potential shape at the selected points. This analysis is conducted via the \texttt{Minimize} algorithm in \texttt{Mathematica} to efficiently locate the minimum within the defined parameter ranges and to assess the function's behavior around it.

\section{Results}
\label{sec:V}

\begin{figure*}
    \centering
    \subfloat[Scalar perturbation related to $\text{Log}_{10}\Upsilon_{010}(\tilde{\zeta}_1)$. \label{fig : scalardensityplots}]{\includegraphics[scale=0.4]{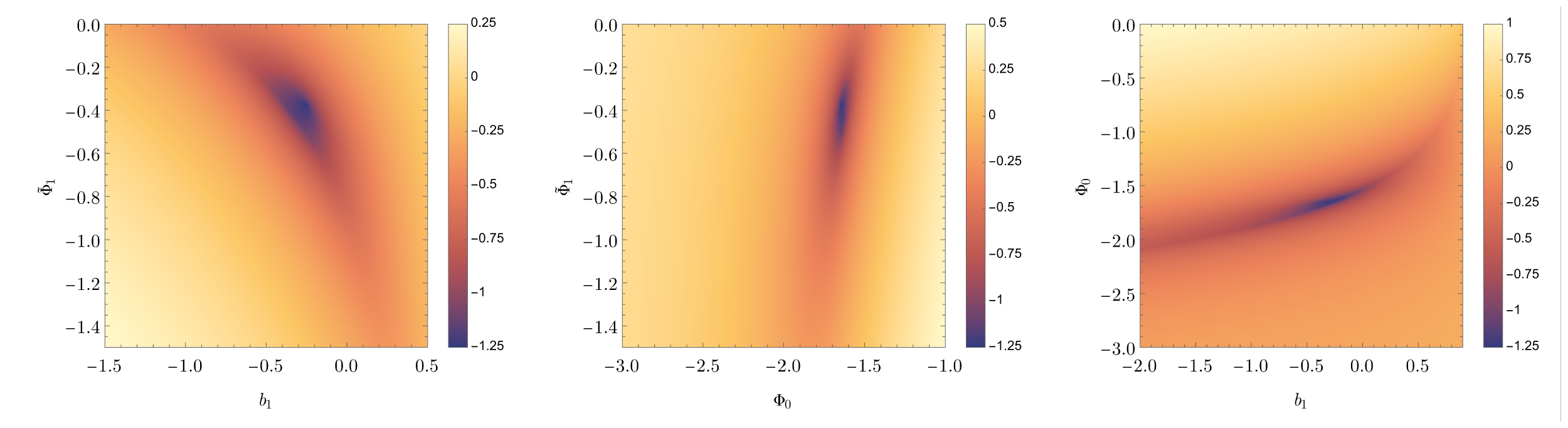}}

    \subfloat[Electromagnetic perturbation related to $\text{Log}_{10}\Upsilon_{111}(\tilde{\zeta}_1)$.\label{fig : emdensityplots}]{\includegraphics[scale=0.4]{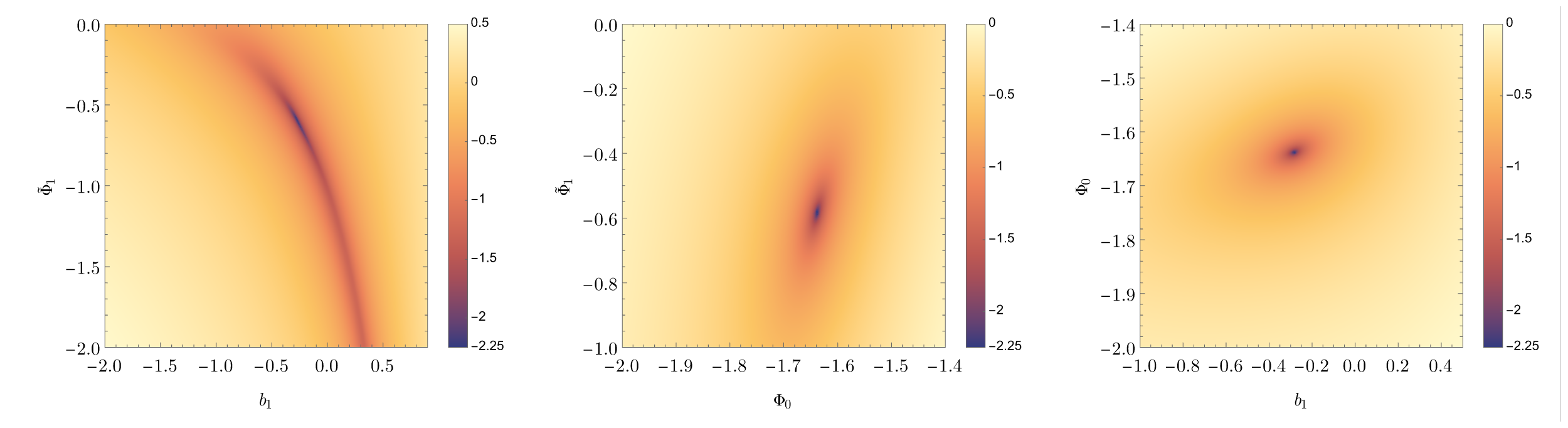}}

    \subfloat[Axial gravitational perturbation related to $\text{Log}_{10}\Upsilon_{122}(\tilde{\zeta}_1)$.\label{fig : axialdensityplots}]{\includegraphics[scale=0.4]{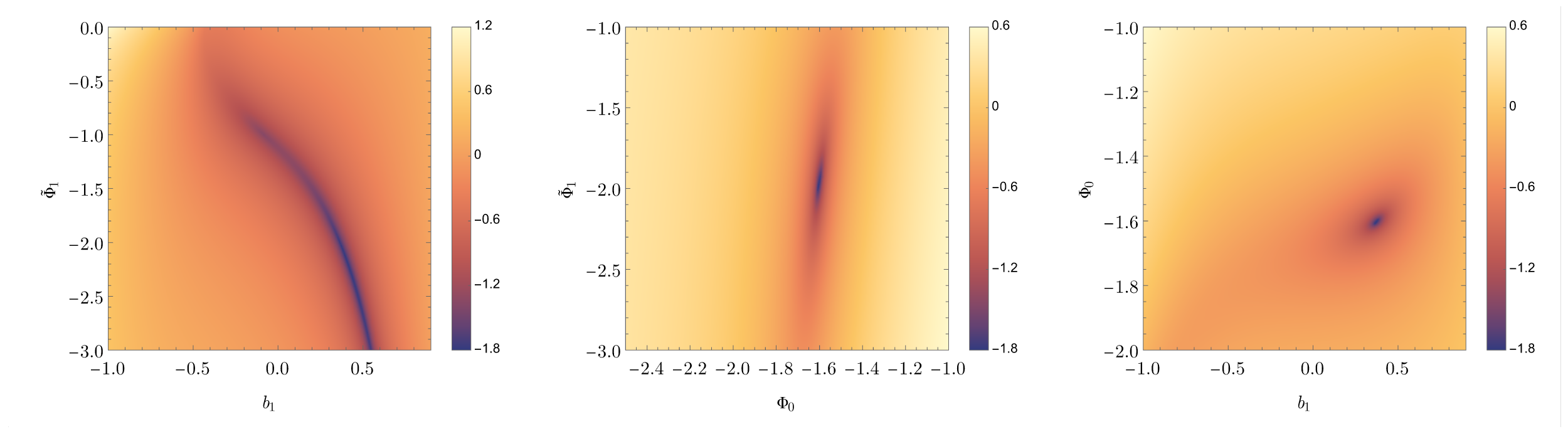}}
    
    \caption{Density plots of the different perturbation types. In all cases, we divide the three-dimensional parameter space spanned by $\tilde{\zeta}_1$ in three two-dimensional plots, being $(b_1,\tilde{\Phi}_1)$, $(\Phi_0,\tilde{\Phi}_1)$, and $(b_1,\Phi_0)$. The lateral bar refers to $\text{Log}_{10}\Upsilon_{\rm nls}(\tilde{\zeta}_1)$ values.}
    \label{fig:Fig1}
\end{figure*}

\begin{figure*}
    \subfloat[Combined perturbations related to $\text{Log}_{10}\bar{\Upsilon}(\tilde{\zeta}_1)$.\label{fig : combineddensityplots}]{\includegraphics[scale=0.4]{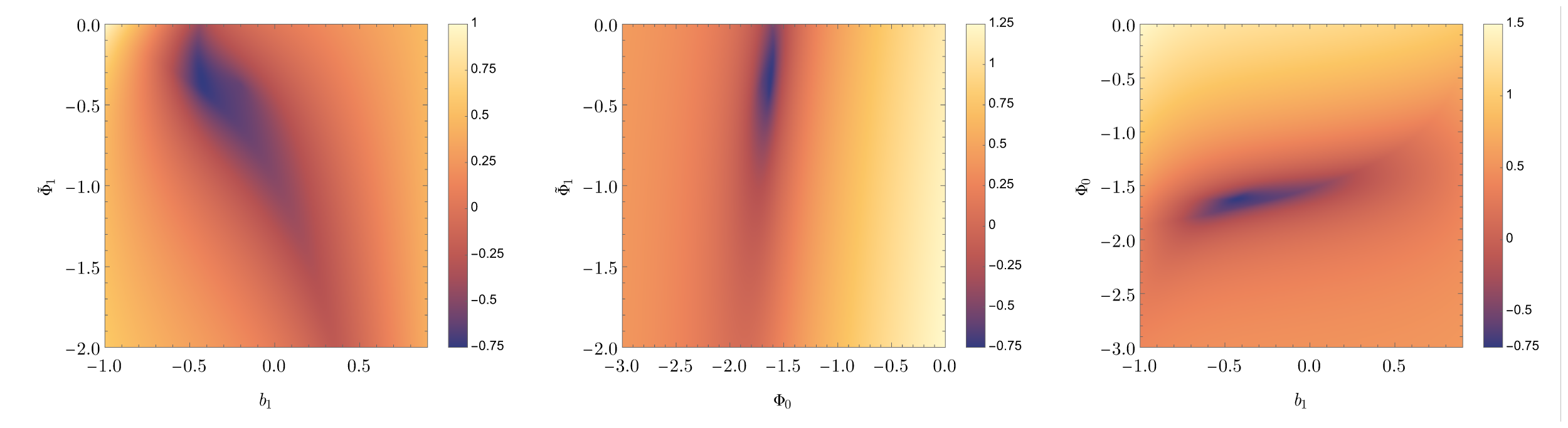}}

    \subfloat[Combined perturbations related to $\text{Log}_{10}\bar{\Upsilon}(\tilde{\zeta}_2)$.\label{fig : combineddensityplots2}]{\includegraphics[scale=0.25]{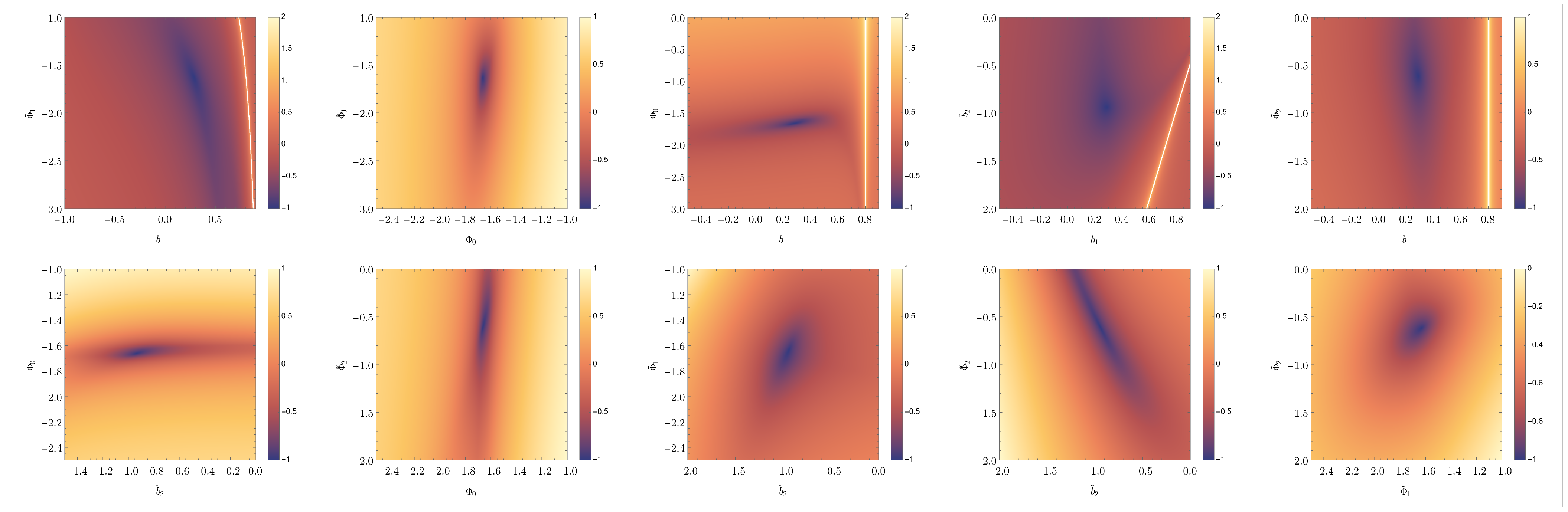}}

    \caption{Density plots of the combined perturbations for the order $N=1$ and $N=2$ of the near-throat parametrization. We split the three- ($N=1$) and five-dimensional ($N=2$) parameter spaces in all possible two-dimensional combinations. The lateral bars in each density plot refers to the $\text{Log}_{10}\bar{\Upsilon}(\tilde{\zeta}_N)$ values.}
    \label{fig: Fig2}
\end{figure*}

For application purposes, we need to truncate the near-throat parametrization \eqref{eq:PARAMETERS} to a certain finite value $N$, implying a $2N+2$ total parameters to be determined. Naturally, increasing $N$ entails a more accurate description of the metric near the WH throat, but the numerical analysis becomes more cumbersome due to more parameters involved. Defining the rescaled radial coordinate $x=r/b_0$, the set of parameters $\zeta_N$ is transformed into
\begin{equation}
\tilde{\zeta}_N=\{b_1,\tilde{b}_2,\tilde{b}_3,\cdots,\tilde{b}_N,\Phi_0,\tilde{\Phi}_1,\tilde{\Phi}_2,\cdots,\tilde{\Phi}_N\},    
\end{equation}
where $\tilde{b}_j=b_j b_0^{j-1}$ for $j \geq 2$ and $\tilde{\Phi}_k=\Phi_k b_0^k$ for $k \geq1$. 
 
In our study, we chose $N=1$, resulting in the three-dimensional space $\tilde{\zeta}_1=\{b_1,\Phi_0,\tilde{\Phi}_1\}$ and representing the minimal set of parameters for the WH model. 

We examine this region of the parameter space
\begin{equation}\label{eq:range}
b_1\in[-5,1), \quad \tilde{\Phi}_0\in[-5,0], \quad \tilde{\Phi}_1\in[-5,0].     
\end{equation}
This selection is primarily motivated by simplicity and can be readily adjusted to explore a broader range of configurations, provided the assumptions underlying the WKB method remain valid. This choice, involving relatively low parameter values, is physically reasonable and is sufficiently wide to include exactly one single minimum. Notice however that, in the case of gravitational perturbations, the compatibility with the WKB method implies the additional constraint $b_1>-2$.

It is important to remember that the WH QNM frequencies $\omega^{\rm WH}_{\rm nls}$ are theoretically expressed in units of the throat radius $b_0$. This implies that, once a mode is observationally measured, $b_0$ can be directly obtained as the ratio between the real (or imaginary) part of observed and theoretical frequencies. In turn, the value of $b_0$ can be used to convert $\tilde{\zeta}_N$ into the original $\zeta_N$.

We analyze $\Upsilon_{\rm nls}(\tilde{\zeta}_1)$ for each individual perturbation type ($s=0,1,2$) and present the results in Fig. \ref{fig:Fig1} as three two-dimensional density plots, depicting all possible parameter pairs. In the study of each singular perturbation, two modes are involved\footnote{We examine the modes $\omega^{\rm WH}_{000}$, $\omega^{\rm WH}_{010}$, $\omega^{\rm WH}_{011}$, $\omega^{\rm WH}_{111}$, $\omega^{\rm WH}_{022}$, and $\omega^{\rm WH}_{122}$ implemented in the different $\Upsilon_{\rm nls}(\tilde{\zeta}_1)$, ensuring they satisfy $n\leq l$ to enable the WKB approximation to function properly.}, yielding thus four independent components when separated into real and imaginary parts. The $\tilde{\zeta}_1$ values that minimize $\Upsilon_{\rm nls}$ are listed in Tab. \ref{tab: parameters minima}. The electromagnetic case exhibits the closest match between WH and BH spectra, as $\Upsilon_{\rm nls}$ attains its minimum value, whereas for scalar and axial gravitational perturbations the agreement remains strong but less pronounced. This is further confirmed by the density plots, where the region of optimal concordance is more confined in the electromagnetic configuration compared to the other two cases. These findings suggest that isospectrality is limited to a narrow region of the parameter space and is highly sensitive to small variations.

Further insights into the inquired isospectrality are obtained by studying $\delta\omega_{\rm nls}^{(\rm r)}(\tilde{\zeta}_1)$ and $\delta\omega_{\rm nls}^{(\rm i)}(\tilde{\zeta}_1)$ at the previously determined minimum $\tilde{\zeta}_1$. This analysis is extended to higher overtones to assess the compatibility beyond the first level. Numerical results for each perturbation type are presented in Tab. \ref{tab: comparison}, where the lowest modes yield the best agreement, but the match remains still robust for higher overtones. This highlights the reliability of the methodology across multiple spectral components.

These findings indicate that for a WH to accurately replicate the QNM spectrum of a BH across various spins and overtones, the order $N=1$ is inadequate, necessitating more detailed knowledge of throat properties. Building on this result, the analysis of the combined case is extended to order $N=2$, expanding the parameter space to the five-dimensional $\tilde{\zeta}_2$. Consistent with the other parameters, the following range is selected:
\begin{equation}\label{eq:range2}
\tilde b_2\in[-5,0],  \quad \tilde{\Phi}_2\in[-5,0].     
\end{equation}
The outcomes of this extended analysis is reported in Fig. \ref{fig : combineddensityplots2}, where we produce ten two-dimensional density plots of all possible combinations of pair parameters. The minimum $\tilde{\zeta}_2$ values are summarized in Table \ref{tab: parameters minima}, where we note that $\bar{\Upsilon}(\tilde{\zeta}_2)$ reduces by approximately $50\%$ with respect to $\bar{\Upsilon}(\tilde{\zeta}_1)$, and is relocated to a slightly different region of the parameter space. However, if the parameters are fixed at the minimum $\bar{\Upsilon}(\tilde{\zeta}_1)$ and only $(b_2, \tilde{\Phi}_2)$ are allowed to vary, the improvement becomes negligible.

As a final remark, we emphasize the importance of ensuring that the perturbation potentials $V^{\rm (s)}_l$ for $s=0,1,2$ exhibit all a single peak at the values $\tilde{\xi}_1$ that minimize $\Upsilon_{\rm nls}(\tilde{\xi}_1)$. The electromagnetic and gravitational scenarios present no issues, while the scalar case is more subtle. Indeed, within certain parameter combinations, the potential can become slightly negative, losing thus its single-peaked structure. Upon closer examination, we observe that $V^{\rm (0)}_l(r_*)$ becomes zero for $r_*\sim(1.7-118.8)b_0$, taking into account the whole intervals of parameter space variation. Since these anomalies in $V^{\rm (0)}_l(r_*)$ arise far from the peak (occurring at $r_*=0$) and we are employing the near-throat parametrization, the WKB method remains applicable, allowing us to display related density plots over the full parameter range. Naturally, the above considerations also extend, with similar caveats, to the case of $\bar{\Upsilon}(\tilde{\xi}_N)$ for both $N=1$ and $N=2$.

\begin{table}[th!]
\renewcommand{\arraystretch}{1.8}
\centering
\caption{Values of $\tilde{\zeta}_1=(b_1,\Phi_0,\tilde{\Phi}_1)$ and $\text{Log}_{10}\Upsilon_{\rm nlm}$ (including $\text{Log}_{10}\bar{\Upsilon}$ for $\tilde{\zeta}_1$ and $\tilde{\zeta}_2$) for different perturbations, obtained from the minimization procedure devised in Sec. \ref{sec:IVB}.\vspace{0.2cm}}
\scalebox{0.95}{
\begin{tabular}{|c|c|c|c|c|c|}
\hline
$\quad\boldsymbol{\Upsilon_{\rm nlm}}\quad$ & $\Upsilon_{010}$ & $\Upsilon_{111}$ & $\Upsilon_{122}$ &  $\bar\Upsilon ( N=1 )$ & $\bar\Upsilon ( N=2 )$\\ \hline
$\boldsymbol{\text{Log}_{10}\Upsilon_{\rm nlm}}$ & -1.3 & -2.3 & -1.8 & -0.7 & -1.0\\ \hline
$\quad\boldsymbol{b_1}\quad$ & $\enspace -0.2512 \enspace$ & -0.2793 & 0.3642 & -0.4266 & 0.3013\\ \hline
$\quad\boldsymbol{\tilde b_2}\quad$ & 0 & 0 & 0 & 0 & -0.9344 \\ \hline
$\quad\boldsymbol{\Phi_0}\quad$ & -1.6472 & -1.6396 & −1.6085 & −1.6264 & -1.6613\\ \hline
$\quad\boldsymbol{\tilde{\Phi}_1}\quad$ & -0.3726 &  -0.5801 & −2.0000 & -0.3470 & -1.6453\\ \hline
$\quad\boldsymbol{\tilde{\Phi}_2}\quad$ & 0 & 0 & 0 & 0 & -0.6328\\ 
 \hline
\end{tabular}}
\label{tab: parameters minima}
\end{table}

\renewcommand{\arraystretch}{1.8}
\begin{table}[!ht]
\centering
\caption{Relative errors of real and imaginary parts pertaining to some QNM overtones related to scalar, electromagnetic, and axial gravitational perturbations. \vspace{0.2cm}}
\begin{tabular}{|c|c|c|} 
\hline
$\qquad\boldsymbol{\omega^{\rm WH}_{\rm nls}}\qquad$ & $\qquad\boldsymbol{\delta\omega^{\rm (r)}_{\rm nls}(\%)}\qquad$ & $\qquad\boldsymbol{\delta\omega^{\rm (i)}_{\rm nls}(\%)}\qquad$\\
\hline
\multicolumn{3}{|c|}{\bf Scalar} \\
\hline
$\omega^{\rm WH}_{000}$ & $\sim 10^{-8}$ & $\sim10^{-8}$\\
\hline
$\omega^{\rm WH}_{010}$ & 0.4 & 5.3 \\
\hline
$\omega^{\rm WH}_{110}$ & 1.5 & 6.1 \\
\hline
$\omega^{\rm WH}_{020}$ & 0.1 & 6.5 \\
\hline
\multicolumn{3}{|c|}{\bf Electromagnetic} \\ 
\hline
$\omega^{\rm WH}_{011}$ & 0.2 & 0.5 \\
\hline
$\omega^{\rm WH}_{111}$ & $\sim10^{-6}$ & $\sim10^{-6}$\\
\hline
$\omega^{\rm WH}_{021}$ & 0.6 & 0.8\\
\hline
$\omega^{\rm WH}_{121}$ & 1.0 & 0.6\\
\hline
\multicolumn{3}{|c|}{\bf Gravitational (axial)} \\
\hline
$\omega^{\rm WH}_{022}$ & 1.6 & 0.1\\
\hline
$\omega^{\rm WH}_{122}$ & $\sim10^{-7}$ & $\sim10^{-8}$ \\
\hline
$\omega^{\rm WH}_{032}$ & 2.1 & 1.9 \\
\hline
$\omega^{\rm WH}_{132}$ & 2.8 & 1.8 \\
\hline
\multicolumn{3}{|c|}{\bf Combination of $s=0, 1, 2$ (N=1)} \\
\hline
$\omega^{\rm WH}_{000}$ & 9.2 & 7.5\\
\hline
$\omega^{\rm WH}_{011}$ & $1.2$ & $2.5$ \\
\hline
$\omega^{\rm WH}_{022}$ & 4.3 & 0.7 \\
\hline
\multicolumn{3}{|c|}{\bf Combination of $s=0, 1, 2$ (N=2)} \\
\hline
$\omega^{\rm WH}_{000}$ & $\sim10^{-7}$ & $\sim10^{-8}$ \\
\hline
$\omega^{\rm WH}_{011}$ & 2.7 & 0.9\\
\hline
$\omega^{\rm WH}_{022}$ & 6.1 &  3.1\\
\hline
\end{tabular}
\label{tab: comparison}
\end{table}

\section{Conclusions}
\label{sec:VI}
In this paper, we have investigated whether a static and spherically symmetric WH solution \eqref{MT metric} in GR is able to replicate the QNM spectrum of a Schwarzschild BH. To address this issue, we have employed several different approaches, but well-connected among them: (1) a near-throat parametrization \eqref{eq:NT-parametrization} to highlight dynamics near the WH throat, where strong gravitational effects dominate; (2) a detailed analysis of scalar, electromagnetic, and axial gravitational perturbations, both individually and collectively; (3) computation of QNM frequencies using the third-order WKB method, which, while slightly less precise than some alternatives, provides significantly greater computational efficiency; (4) a tailored methodology for systematically comparing BH and WH spectra in GR over an extensive range of metric parameters.

To carry out the calculations more rapidly, we employ the minimal parameter configuration permitted by the near-throat parametrization, specifically $N=1$. \emph{We can say that a WH can effectively emulate the Schwarzschild BH in GR in its fundamental mode and first overtone across the three distinct perturbation types considered individually}. Furthermore, when analyzing the fundamental modes encompassing all three perturbation types simultaneously, the WH still reproduces BH characteristics, albeit with reduced precision. To address this limitation, we extended, only for this case, the analysis to $N=2$, which yielded improved agreement. These findings align with previous studies \cite{DamourWH, KonoplyaWH} and advance them by incorporating perturbations with various spin, as well as with a more systematic and rigorous analysis.

This study offers several avenues for further and different investigations. A natural extension involves refining the near-throat parametrization by advancing beyond the first-order approximation ($N>1$), trying to develop a strategy to drastically reduce the computational times. This enhancement is advantageous, as it permits to: (i) have a more precise representation of the WH throat's properties (particularly significant for the combination of QNM modes with different spin, as we have already seen in Sec. \ref{sec:IV}); (ii) include higher overtones for extending the numerical isospectrality to a broader set of QNMs; (iii) extract more accurately the model coefficients $\zeta_N$ from a specific WH metric. Finally, the metric parameters can serve also as an approach to explore the phenomenology of a WH in other physical regimes, e.g., the X-ray electromagnetic sector, to assess whether these similarities with the Schwarzschild BH persist \cite{Volkel2021,Delijski2022}. Such analyses, however, require the full explicit expressions of the metric coefficients, including also their pre-factors. Those functions can be chosen independently from the near-throat parametrization, as long as they do not affect the behavior of the metric near the throat. From this perspective, the determination of the pre-factors is an interesting possible development of this work. This encourages also to develop alternative strategies or other approaches to solve these thorny issues.

As a near-term objective, we plan to extend this study to polar gravitational perturbations to enable a more comprehensive comparison between the QNMs of BHs and WHs in GR. However, polar modes typically necessitate accounting for variations in  pressure and density of the object, which, in turn, requires specifying the WH stress-energy tensor. In cases where no specific fluid model is assumed, the stress-energy tensor must be represented in a parametrized form, inevitably increasing the number of free parameters in the ensuing analysis. Therefore, we need to develop a strategy able to solve the problem and overcome the computational difficulties.

\section*{Acknowledgements}
V.D.F. and S.C. thank the  Gruppo Nazionale di Fisica Matematica (GNFM) of Istituto Nazionale di Alta Matematica (INDAM) for the support. C.D.S., V.D.F., and S.C. acknowledge the support of INFN {\it sez. di Napoli}, {\it iniziative specifiche} QGSKY, TEONGRAV, and MOONLIGHT2. This paper is based upon work from COST Action CA21136 - Addressing observational tensions in cosmology with systematics and fundamental physics (CosmoVerse), supported by COST (European Cooperation in Science and Technology). The authors are very grateful to Dr. Sebastian V{\"o}lkel for the useful comments and discussions he provided us.

\appendix
\section{Derivation of the potential related to axial gravitational perturbations}
\label{AppendixA}

In order to study the axial gravitational perturbations over the WH spacetime \eqref{MT metric}, we generalize the results obtained for the Schwarzschild BH in Ref. \cite{Chandrasekhar1998book}. The line element for the perturbations of a WH can be written as:
    \begin{align}
		\dd s^2& =-e^{ 2\Lambda(r)}\dd t^2+\frac{1}{1-\frac{b(r)}{r}}\dd r^2+r^2 \dd\theta^2+r^2 \sin^2\theta\,\Big{[}\dd\varphi\nonumber\\ 
        &-q_1(r,\theta) \dd t-q_2(r,\theta) \dd r-q_3(r,\theta) \dd\theta\Big{]}^2,
	\end{align}
where $q_1(r,\theta),q_2(r,\theta),q_3(r,\theta)$ are arbitrary functions.

As mentioned in Sec. \ref{sec:IIIC}, axial perturbations induce a rotation of the spacetime and are thus characterized by a nonzero variation of $q_1(r,\theta),q_2(r,\theta)$, and $q_3(r,\theta)$, while the other coefficients do not vary. Since axial gravitational perturbations are not affected by the perturbations of the matter content, by introducing the functions
    \begin{equation}
        Q(r,\theta) = r \,e^{\Lambda(r)}\sqrt{\Delta(r)} Q_{23}(r,\theta)\sin^3\theta,
    \end{equation}
where $\Delta(r)=r^2-r\,b(r)$ (cf. under Eq. \eqref{eq:potential_axial}) and 
\begin{align}
Q_{23}(r,\theta)&=\frac{\partial
q_{2}(r,\theta)}{\partial\theta}-\frac{\partial q_{3}(r,\theta)}{\partial r},
\end{align}    
they will be governed by the Einstein field equations $\delta R_{12}=\delta R_{13}=0$, which reads as \cite{Chandrasekhar1998book}
\begin{subequations}
    \begin{align}
        \frac{e^{\Lambda}}{r^3\sqrt{\Delta}\sin^3\theta}\frac{\partial Q}{\partial\theta}&=i\omega \frac{\partial q_1}{\partial r}-\omega^2 q_2, \label{eq: pert1}\\
        \frac{e^{\Lambda}\sqrt{\Delta}}{r^3\sin^3\theta} \frac{\partial Q}{\partial r}&=-i\omega \frac{\partial q_1}{\partial \theta}+\omega^2q_3 .\label{eq: pert2}
    \end{align}
\end{subequations}    
Summing the derivative of Eq. (\ref{eq: pert1}) with respect to $\theta$ and the derivative of Eq. (\ref{eq: pert2}) with respect to $r$, the terms in $q_1$ cancel and we have an equation only in $Q$:
    \begin{align}\label{eq: axialpert}
        &\frac{e^{\Lambda}}{r^3\sqrt{\Delta}}\sin^3\theta \frac{\partial}{\partial \theta}\left(\frac{1}{\sin^3 \theta}\frac{\partial Q}{\partial \theta}\right)+\frac{\omega^2 Q}{r e^{\Lambda}\sqrt{\Delta}} \nonumber\\
        &+\frac{\partial}{\partial r}\left( \frac{e^{\Lambda}\sqrt{\Delta}}{r^3}\frac{\partial Q}{\partial r}\right)=0.
    \end{align}
This equation is solved for $Q(r,\theta)=\mathcal{Q}(r)\,\mathcal{C}_{l+2}^{-3/2}(\theta)$, where $\mathcal{C}_{l+2}^{-3/2}(\theta)$ is a specific instance of the Gegenbauer function $\mathcal{C}_{n}^{\nu}(\theta)$ solving the following differential equation
    \begin{equation}
        \left[\frac{\dd}{\dd\theta} \sin^{2\nu}\theta\frac{\dd}{\dd\theta}+n(n+2\nu)\sin^{2\nu}\theta\right]\mathcal{C}_{n}^{\nu}(\theta)=0.
    \end{equation}
Therefore, after the substitution of $Q(r,\theta)$, Eq. \eqref{eq: axialpert} becomes only an equation for $\mathcal{Q}(r)$. By introducing the function $Z(r)=\mathcal{Q}(r)/r$ and the tortoise coordinate \eqref{eq:tortoise_coordinate}, Eq. \eqref{eq: axialpert} acquires the Schr{\"o}dinger-like form 
	\begin{equation}\label{eq:final-AG}
		\frac{\dd^2 Z(r)}{\dd r_*^2}+\Big{[}\omega^2-V_l^{(2), \rm axial}(r)\Big{]}Z(r)=0,
	\end{equation} 
with the potential $V_l^{(2), \rm axial}(r)$ reported in Eq. \eqref{eq:potential_axial}. A similar derivation can be found also in Ref. \cite{Kim2008}. Although the final expression for the potential is correct, there is a typo in the definition of $Q(r,\theta)$. It is possible to check that Eq. \eqref{eq:final-AG} reduces to the Regge-Wheeler equation if we substitute the Schwarzschild BH solution.

\bibliography{references}

\end{document}